\documentclass[journal,article,accept,oneauthor,pdflatex,12pt,a4paper]{mdpi} 
\lastpage{x}
\doinum{10.3390/------}
\pubvolume{17}
\pubyear{2015}
\externaleditor{Academic Editor: Remo Garattini}
\history{Received: 31 July 2015/ Accepted: 8 September 2015/ Published: xx September 2015}
\usepackage{amsmath}
\usepackage{amssymb}
\usepackage{soul}

\Title{Viscosity-Induced Crossing of the Phantom Barrier}

\Author{Iver Brevik}

\address[1]{%
Department of Energy and Process Engineering, Norwegian University of Science and Technology, N-7491 Trondheim, Norway; E-Mail:~iver.h.brevik@ntnu.no; Tel.:~+47-7359-3555; Fax:~+47-7359-3491\vspace{12pt}
}

\corres{}

\abstract{We show explicitly, by using astrophysical data plus reasonable assumptions for the bulk viscosity in the cosmic fluid, how the magnitude of this viscosity may be  high enough  to drive the fluid  from  its  position in the  quintessence region at present time \linebreak $t=0$~across the barrier $w=-1$ into the phantom region in the late universe. The phantom barrier is accordingly not a sharp mathematical divide, but rather a fuzzy concept. We also calculate the limiting forms of various thermodynamical quantities, including the rate of entropy production,
for a dark energy fluid near the future Big Rip singularity.}


\begin{document}

\section{Introduction} \label{s1}

Recent years have seen an increased interest in viscous cosmology. This contrasts the traditional approach, in which  the cosmic fluid has usually be taken to be ideal (nonviscous).  From a hydrodynamicist's point of view this is to some extent  surprising, since there are several situations in fluid mechanics, even in homogeneous space without boundaries,  where the inclusion of viscosity becomes mandatory. Especially, this is so in connection with turbulence phenomena. As is known, to the first order in deviation from thermal equilibrium there are two such coefficients, the shear viscosity $\eta$, most often being the dominant one, and the bulk viscosity $\zeta$ \cite{landau87}. Because of the assumption about spatial isotropy in cosmology, the shear coefficient is usually omitted. (As a parenthetical remark we note that this first-order theory away from thermal equilibrium means in effect acceptance of  the Eckart 1940 theory \cite{eckart40}, implying noncausality.   By taking into account second order deviations from thermal equilibrium one can, however,  obtain a causal theory respecting special relativity. Pioneering articles on causal fluid mechanics are those of   M{\"u}ller \cite{muller67}, Israel \cite{israel76}, and Israel and Stewart \cite{israel79}.)

Viscosity coefficients have so far been considered before, both for the past ($t<0$) and for the future ($t>0$) universe. Since a large amount of knowledge is available about the past universe, it is natural to apply methods from particle physics in order to understand the viscosity's influence upon the evolution. Thus, Hogeveen \emph{et al.} \cite{hogeveen86} give results for both shear and bulk viscosity in the very early universe, calculated from kinetic theory. The general tendency seems to be that the modifications from viscosity to the development of the early universe are moderate or small. In the future universe, however, as we will focus on below, much less is known, and it becomes therefore natural to make use of phenomenological suggestions for the magnitude of the viscosity. Some articles and reviews dealing with viscous cosmology from a phenomenological point of view are Refs.~\cite{weinberg71,gron90,zimdahl96,cardone06,mostafapoor11,brevikgron13}.

We will in the following consider the future evolution of the cosmic fluid, where for simplicity this is taken to consist of  one component only. This is the dark energy fluid endowed with an equation-of-state parameter $w$. Again for simplicity, and also because of  the desirability to keep the number of input parameters in the theory restricted, we shall assume that $w$ is a constant. Thus, taking the equation of state to be homogeneous, we can write
\begin{equation}
p=w\rho, \quad w=-1+\alpha. \label{1}
\end{equation}
Astronomical observations show that $w$ is close to unity, thus, $\alpha$ very small. According to the 2015  Planck data, $w=-1.019^{+0.075}_{-0.080}$ \cite{ade15} (see Table 5). The region $-1<w<-1/3$ is the quintessence region, while $w<-1$ is the phantom region. The case $w=-1$ is called the phantom barrier.

In this paper we will deal with the phantom region and the phantom barrier. For some years it has been known that if the cosmic fluid starts out from some value of $w$ lying in the phantom region, it will encounter some form of singularity in the remote future. The most dramatic event is called the Big Rip, in which the fluid enters into a singularity after a finite time span \cite{caldwell03,nojiri03,nojiri04}. There are also softer variants of the future singularity where the singularity is not reached until an infinite time, called the Little Rip~\cite{frampton11,brevik11,brevik12},  the Pseudo Rip \cite{frampton12}, and the Quasi Rip \cite{wei12}.

In view of the mentioned smallness of $|\alpha|$ it becomes natural to ask: if the cosmic fluid starts from present time $t=0$ in the quintessence region ($\alpha>0$), has it any possibility to slide through the phantom barrier into the phantom era and thereafter inevitably be drawn into the future singularity? This problem was  analyzed already in 2005 \cite{brevik05} with the following result: if the magnitude of the bulk viscosity is large enough, and if the viscosity varies with the energy in a  way that is natural physically, such a transition is actually possible. The analysis in Ref.~\cite{brevik05} was however restricted to qualitative considerations, as quantitative  information about the viscosity was not available. Recently this situation has improved,  due to the analysis of Wang and Meng \cite{wang14} and others on the influence from viscosity on the Friedmannn equations giving the evolution of  the Hubble parameter, $H=H(z)$, $z$ being the redshift. Introduction of the quantitative element  is the main point in the present study. As will be shown, reasonable values of the viscosity inferred from the experiments are just of the right order of magnitude necessary to make the transition through the phantom barrier possible.

 It has recently come to our attention that there exists  a recent series of papers by Velten \emph{et al.}  dealing with viscous cosmology \cite{velten12,velten13,velten13a,velten14}, of definite interest for our work. We shall have the opportunity to compare with some of those results later. See also the related  paper of Nunes and Pav\'{o}n \cite{nunes15}.

In the next section we present the general formalism, list the various actual forms for the relationship $\zeta = \zeta(\rho)$, and discuss the fluid's possibilities for phantom barrier passage. In Section \ref{s3} we consider further comparisons with experiments. In Section \ref{s4} we discuss, under simplifying assumptions,  limiting forms for various thermodynamic quantities near the future singularity.

\section{Basic Formalism. Various Assumptions for the Bulk Viscosity}\label{s2}

We first have to sketch the basic formalism as outlined, for instance, in Refs.~\cite{brevik05} and \cite{brevik10}. We start from the standard FRW metric
\begin{equation}
ds^2=-dt^2+a^2(t)d{\bf x}^2, \label{2}
\end{equation}
when the  spatial curvature $k$ is set equal to zero.

The Friedmann equations, with $\theta=3H=3\dot{a}/a$ the scalar expansion, are
\begin{equation}
\theta^2=24\pi G\left( \rho +\frac{\Lambda}{8\pi G}\right), \label{3}
\end{equation}
\begin{equation}
\dot{\theta}+\frac{1}{2}\theta^2=-12\pi G\left(p-\zeta(\rho)\theta-\frac{\Lambda}{8\pi G}\right), \label{4}
\end{equation}
and the energy conservation equation is
\begin{equation}
\dot{\rho}+(\rho+p)\theta =\zeta(\rho)\theta^2. \label{5}
\end{equation}
We set henceforth the cosmological constant $\Lambda=0$. The  governing equation for the scalar expansion then becomes
\begin{equation}
\dot{\theta}+\frac{1}{2}\alpha \theta^2-12\pi G \zeta(\rho)\theta=0, \label{6}
\end{equation}
which can in view of Equation~(\ref{3}) be rewritten as
\begin{equation}
\dot{\rho}+\sqrt{24\pi G}\,\alpha \rho^{3/2}-24\pi G\zeta(\rho)\rho=0. \label{7}
\end{equation}
The solution is
\begin{equation}
t=\frac{1}{\sqrt{ 24\pi G}}\int_{\rho_0}^\rho \frac{d\rho}{\rho^{3/2}\left[ -\alpha +\sqrt{24\pi G}\,\zeta(\rho)/{\sqrt \rho}\right]}, \label{8}
\end{equation}
where we integrate from present time $t=0$  (with density  $\rho_0$), into the future. Note that the sign of $\alpha$ as defined in Equation~(\ref{1}) is the opposite of that used in Ref.~\cite{brevik05}.

We will now analyze different options for $\zeta(\rho)$.

\subsection{$\zeta$ Equal to a Constant}\label{s21}

In this case it is most convenient to go back to Equation~(\ref{6}) and solve directly for $\theta$:
\begin{equation}
\theta(t)=\frac{\theta_0 e^{t/t_c}}{1+\frac{1}{2}\alpha \theta_0t_c(e^{t/t_c}-1)}, \label{9}
\end{equation}
where $t_c=(12\pi G\zeta)^{-1}$ is the ``viscosity time''.

Correspondingly, we obtain for the scale factor
\begin{equation}
a(t)=a_0\left[ 1+\frac{1}{2}\alpha \theta_0t_c(e^{t/t_c}-1)\right]^{\frac{2}{3\alpha}}, \label{10}
\end{equation}
and for the density
\begin{equation}
\rho(t)=\frac{\rho_0e^{2t/t_c}}{\left[ 1+\frac{1}{2}\alpha \theta_0t_c(e^{t/t_c}-1)\right]^2}. \label{11}
\end{equation}
We thus see that if the universe starts out from the phantom region $(\alpha<0)$, it will inevitably be driven into the Big Rip singularity at a well-defined rip time $t=t_s$, where
\begin{equation}
t_s=t_c\ln \left( 1+\frac{2}{|\alpha|\theta_0t_c}\right). \label{12}
\end{equation}
Both $\theta(t), a(t),$ and $\rho(t)$ diverge at this instant. However, if it starts out in the quintessence region $(\alpha>0)$, the behavior is quite different: we see that in the remote future
\begin{equation}
\theta(t)\rightarrow \frac{2}{\alpha t_c}, \quad t\rightarrow \infty, \label{13}
\end{equation}
\begin{equation}
a(t)\rightarrow \infty, \quad t\rightarrow \infty, \label{14}
\end{equation}
\begin{equation}
\rho(t)\rightarrow \frac{4\rho_0}{(\alpha \theta_0t_c)^2}, \quad t\rightarrow \infty. \label{15}
\end{equation}
Observe that the finite limiting values for $\theta(t)$ and $\rho(t)$ are viscosity dependent. For an ideal fluid $(t_c\rightarrow \infty)$, these two quantities go to zero when $t\rightarrow \infty$.

We see that in this model the case $\alpha=0$ acts as a sharp boundary; the behavior of the fluid when $\alpha$ is positive is very different from the behavior when $\alpha$ is negative. This indicates that  the assumption about constant $\zeta$ is over-idealized; a real physical fluid is not expected to behave in such a singular way, at least not so in a classical theory. We therefore turn to models which are physically more plausible. There are two models of that kind, to  be considered successively below.

\subsection{$\zeta$  Proportional to $\theta$}

In the late universe, especially near the future singularity where the fluid motion is vigorous and the occurrence of turbulence highly likely, it is natural to assume that the bulk viscosity increases with respect to time. We consider first a model where $\zeta \propto \theta$,
\begin{equation}
\zeta(\rho)=\tau_1 \theta =\tau_1 \sqrt{24\pi G \rho}, \label{16}
\end{equation}
with $\tau_1$ a constant. Then Equation~(\ref{8}) yields
\begin{equation}
t=\frac{1}{\sqrt{24\pi G}}\,\frac{2}{-\alpha+24\pi G\tau_1}\left(\frac{1}{\sqrt{\rho_0}}-\frac{1}{\sqrt \rho}\right), \label{17}
\end{equation}
with $\rho_0$ referring to the present time $t=0$. If the universe starts out from the phantom region, the development into a Big Rip singularity is inevitable. But if it starts out from the quintessence region, the development depends critically on   how large the coefficient $\tau_1$ is. The condition for passing through the phantom barrier and into the phantom region is seen to be
\begin{equation}
\frac{24\pi G\tau_1}{c^2} >\alpha, \label{18}
\end{equation}
when expressed in dimensional units.
If this condition holds, the Big Rip $(\rho=\infty)$ is reached in a finite~time.

This possibility, pointed out already in Ref.~\cite{brevik05}, can now be supplemented by data inferred from experiments, though only on the order-of-magnitude level. We shall first consider the analysis of Wang and Meng \cite{wang14}. They include various assumptions for the bulk viscosity in the early universe in the Friedmann equations and compare the theoretical curve for $H=H(z)$ with a number of observations. The comparison gets somewhat complicated, due to the various parameters involved, and also because of allowance of a more general equation of state than the form $w=-1+\alpha$ which we have  chosen above for simplicity. We intend to analyze this issue more closely in a forthcoming paper \cite{normann15}. For our present purpose the relevant information can however be drawn relatively easily: a central quantity is  $B$, which we shall define  as
\begin{equation}
B=\frac{12\pi G\zeta_0}{c^2}. \label{19}
\end{equation}
Here $12\pi G/c^2= 2.79\times 10^{-26}$ m/kg, and the dimension of $B$  is $[B]={\rm s}^{-1}$. Again, in $\zeta_0$ the subscript zero refers to present time $t=0$.

From Refs.~\cite{wang14} and \cite{normann15} we infer that, given our simple form for the equation of state, the best agreement with observational data is obtained when
\begin{equation}
|B| \sim 50\,{\rm \frac{km}{s\, Mpc}} =1.62\times 10^{-18}{\rm s}^{-1}. \label{20}
\end{equation}
[Note: there is an apparent difficulty here because the values of $B$ turn out to be negative. This may be due to two causes; either that (i) the bulk viscosity introduced in Friedmann's equations is of a formal, rather than a physical, nature; (ii) or that this behavior is related to our choice for equation of state. We think the latter option is true here. The analysis turns out to be quite sensitive to the different equations of state one adopts for the fluid. More complicated forms for the equation of state easily lead to positive values for $B$. In any case, the magnitudes for $|B|$ turn out to be roughly in agreement with Equation~(\ref{20}). This is sufficient for our purpose.]

From Equations~(\ref{19}) and (\ref{20}) we estimate, on basis of the given information,
\begin{equation}
\zeta_0 \sim 5\times 10^7~ {\rm Pa~ s}. \label{21}
\end{equation}
This value is probably somewhat high. We may compare with the formula given by Weinberg \cite{weinberg71} for the bulk viscosity in a photonic fluid,
\begin{equation}
\zeta = 4a_{\rm rad}T^4\tau_f\left[ \frac{1}{3}-\left( \frac{\partial p}{\partial \rho}\right)_n\right]^2, \label{22}
\end{equation}
where $a_{\rm rad}=\pi^2 k_B^4/15\hbar^3c^3$ is the radiation constant and $\tau_f$ the mean free time. Considering the microwave background at 3 K slightly out of equilibrium with pressure-free matter and estimating the free time to be the inverse Hubble radius, $\tau_f = 1/H_0$, we obtain $\zeta \sim 10^4~$Pa s. Let us in the following simply assume, as a conservative estimate,  that $\zeta_0$ lies in the region
\begin{equation}
10^4~{\rm Pa~s} <\zeta_0 < 10^6~{\rm Pa~s}. \label{23}
\end{equation}
Now go back to Equation~(\ref{18}) and insert for $\tau_1$ present-time quantities, $\tau_1=\zeta_0/\theta_0$, with $\theta_0=3H_0$. The condition for phantom barrier passage can then be written
\begin{equation}
\zeta_0 > \frac{\theta_0c^2}{24\pi G}\,\alpha. \label{24}
\end{equation}
With $H_0=67.80$ km s$^{-1}$ Mpc$^{-1}=2.20\times 10^{-18}~{\rm s}^{-1}$ we obtain
\begin{equation}
\zeta_0> (1.18\times 10^8) \,\alpha. \label{25}
\end{equation}
We gave above (after Equation~(\ref{1}) the value of $w$ according to the 2015 Planck data; this yields the maximum value to be $w_{\rm max}=-1.019+0.075=-0.944$ or $\alpha_{\rm max}=0.056$. In this case, the condition for driving the fluid across the phantom barrier is
\begin{equation}
\zeta_0 > \frac{\theta_0c^2}{24\pi G}\,\alpha_{\rm max}=  6.6\times 10^6~\rm Pa\, s.\label{26}
\end{equation}
 This leads us to the important conclusion that the  the estimates for the present viscosity given in Equation~(\ref{23}) are roughly high enough   to drive the fluid through the phantom barrier, even in the most extreme quintessential case  $w=w_{\rm max}$.

\subsection{$\zeta$ Proportional to $\theta^2$}

Another option would be to assume that the bulk viscosity varies with $\theta$ as
\begin{equation}
\zeta(\rho)=\tau_2\,\theta^2=\tau_2\cdot  24\pi G \rho, \label{27}
\end{equation}
with $\tau_2$ a new constant. This amounts physically to attaching more weight to the stages where the fluid density is high. In this case we obtain, instead of Equation~(\ref{17}) \cite{frontiers},
\begin{align}
 t&=\frac{-2}{\sqrt{24\pi G}}\int_{\sqrt{\rho_0}}^{\sqrt \rho}\frac{ dx}{x^2[\alpha-(24\pi G)^{3/2}\tau_2\,x]}  \notag\\
&=\frac{2}{\sqrt{24\pi G}}\Large\{ \frac{1}{\alpha}\left( \frac{1}{\sqrt \rho}-\frac{1}{\sqrt{\rho_0}}\right)+\frac{(24\pi G)^{3/2}\tau_2}{\alpha^2}\ln \left( \frac{\sqrt {\rho_0}}{\sqrt \rho}\frac{\alpha-(24\pi G)^{3/2}\tau_2\sqrt \rho}{\alpha -(24\pi G)^{3/2}\tau_2 \sqrt{\rho_0}} \right) \Large\}. \label{28}
\end{align}
Again, we find that the future development of the universe is critically dependent on the value of the equation-of-state parameter, assumed constant in the present model. If the initial stage ($t=0$) admits one to put $\alpha <0$, then a future Big Rip singularity will be encountered after a finite time
\begin{equation}
t_s=\frac{2}{|\alpha|}\frac{1}{\sqrt{24\pi G \rho_0}}. \label{29}
\end{equation}
However, if the universe starts from the quintessence region at $t=0$, the expression (\ref{27}) shows a complicated behavior indicating that the expression as such loses physical interest. The reason for this is  that the  assumption $w=-1+\alpha$ that we have made in this paper for simplicity is incompatible with the viscosity ansatz  (\ref{27}) above. One may easily change the equation of state into a more general form $w=w(\rho)$, and in that way maintain the same sliding property of the cosmic fluid \cite{frontiers}.

\section{Further Comparisons} \label{s3}

In our  numerical considerations above, we focused on the analysis of Wang and Meng \cite{wang14}. It is desirable to compare with other investigations also. As mentioned at the end of Section \ref{s1} there are several recent papers in this direction. A paper of great interest is that of Velten \emph{et al.} on dark matter dissipation~\cite{velten12}. A main outcome of their analysis of current data from supernovae, baryon acoustic oscillations and cosmic microwave background, was that dark matter is allowed to have a bulk viscosity $\leq 10^7~$Pa s. This was found to be consistent also with the integrated Sachs--Wolfe effect (which sometimes is a difficult point for  viscous cosmology theories). The agreement with our estimate in Equation~(\ref{23})  is actually seen to be quite good.

It should be mentioned here for the sake of comparison that the nondimensional quantity $\tilde{\xi}$ in Ref.~\cite{velten12} is defined to be essentially  the same as our $B/H_0$,
\begin{equation}
\tilde{\xi}=\frac{2B}{H_0}.
\end{equation}
The authors represent the bulk viscosity in the form
\begin{equation}
\zeta=\zeta_0\left( \frac{\rho}{\rho_0}\right)^\nu,
\end{equation}
with emphasis on the cases $\nu=0$ and $\nu=-1/2$.

Another noteworthy point in connection with Ref.~\cite{velten12} is that the viscous theory was found to be able to form  galactic dark halos (within the hierarchial scenario) only if $\tilde{\xi} \ll 0.2.$ This means
\begin{equation}
\zeta_0 \ll 7.9\times 10^6~\rm{Pa~s}.
\end{equation}
This behavior should accordingly be quite compatible with the range given in Equation~(\ref{23}). Most of the cases shown in Figure~2 in Ref.~\cite{velten12}  are however given for lower viscosities. As an example, the typical value of $\tilde{\xi}=2\times 10^{-4}$ given for galactic clusters corresponds to $\zeta_0=7.9\times 10^3~$Pa s. A general tendency is that the higher viscosities belong to the larger scales (galaxy clusters).

Similar results are obtained in Ref.~\cite{velten14} on structure formation in viscous universes. For instance, if dark matter has a viscosity of $\tilde{\xi}=10^{-5}$, a strong growth suppression is observed at small scales.

\section{Remark on Entropy Production. Behavior of Thermodynamic Quantities near the Future Singularity.} \label{s4}

We may start by applying the standard formula for
   the rate of entropy produced per particle (subscript zero again referring to present time),
\begin{equation}
\dot{\sigma}_0=\frac{\zeta_0}{n_0k_BT_0}\,\theta_0^2. \label{30}
\end{equation}
We may interpret the universe as a ``gas'', letting a ``particle'' be a galaxy,  and we  may insert for the  number density of galaxies the value \cite{lotz11}
\begin{equation}
n \sim 10^{-69}~{\rm m}^{-3}. \label{31}
\end{equation}
Inserting a typical
  bulk viscosity of  $\zeta_0 \sim 10^5~$Pa s, and using  $\theta_0=3H_ 0=6.60\times 10^{-18}~{\rm s}^{-1}$ together with  $T_0=3~$K, we then estimate from Equation~(\ref{30}) an enormous value for the entropy growth which escapes a natural physical interpretation. This may be related to the difficulty of assigning a viscosity to a gas of galaxies. We find it    better to leave this point, and  analyze instead the behavior of the thermodynamic quantities near the future Big Rip, which is a topiof definite physical interest. Let us assume that  the universe starts out with a value of $\alpha$ that is small, but nevertheless negative. For simplicity we will  limit ourselves to the case of constant viscosity, as considered in Section \ref{s21}. Thus, we assume henceforth
\begin{equation}
w=-1+\alpha={\rm constant}, \quad \zeta = \rm constant. \label{33}
\end{equation}
From Equations~(\ref{9})--(\ref{11}) we obtain
\begin{equation}
\theta(t) \rightarrow \frac{2}{|\alpha|}\frac{1}{t_s-t}, \quad t\rightarrow t_s, \label{34}
\end{equation}
\begin{equation}
a(t)\rightarrow \left( \frac{t_c}{1+\frac{1}{2}|\alpha|\theta_0t_c}\right)^{\frac{2}{3|\alpha1}}\, \frac{a_0}{(t_s-t)^{2/3|\alpha|}}, \quad t\rightarrow t_s, \label{35}
\end{equation}
\begin{equation}
\rho(t)\rightarrow \frac{4}{(\alpha \theta_0)^2}\frac{\rho_0}{(t_s-t)^2}, \quad t\rightarrow t_s; \label{36}
\end{equation}
(recall Equation~(\ref{12}) for the rip time $t_s$).

We need also the corresponding limit for $n(t)$. From the continuity equation ${(nU^\mu)}_{;\mu}=0$, we have in the local rest frame $\dot{n}=-n\theta$, which implies
\begin{equation}
n(t)\rightarrow n_0\left( \frac{1+\frac{1}{2}|\alpha|\theta_0 t_c}{t_c}\right)^{\frac{2}{|\alpha|}}(t_s-t)^{\frac{2}{|\alpha|}}, \quad t\rightarrow t_s. \label{37}
\end{equation}
This expression actually goes to zero, at the singularity.

Finally, we  shall find the time dependence of $T(t)$ near the singularity, under the assumption that the viscosity is ``small''. This is taken to mean that the approximation
\begin{equation}
|\alpha|\theta_0t_c \gg 1 \label{38}
\end{equation}
holds. We may actually check this in the present case, by taking $\zeta=10^5~$Pa s as a typical value. In dimensional units we find
\begin{equation}
\theta_0 t_c=\frac{\theta_0 c^2}{12\pi G\zeta}=2.36\times 10^3. \label{39}
\end{equation}
Thus, as long as $|\alpha|$ is of order $10^{-2}$ or less, the approximation (\ref{38}) holds reasonably well. In view of Equation~(\ref{12}), this means that
\begin{equation}
t_s \ll t_c. \label{40}
\end{equation}
We can then find $T(t)$ from the standard formula for nonviscous flow, using $|\alpha| \ll 1$,
\begin{equation}
T(t)=T_0\left( \frac{a}{a_0}\right)^{3(|\alpha|+1)} \approx T_0\left( \frac{a}{a_0}\right)^3 \propto \frac{T_0}{(t_s-t)^{2/|\alpha|}}, \quad t\rightarrow t_s. \label{41}
\end{equation}
Making use of these expressions, we find for the rate  of entropy production per galaxy
\begin{equation}
\dot{\sigma}(t)=\frac{\zeta}{n(t)k_BT(t)}\theta^2(t) \sim  \frac{\zeta}{\alpha^2}\frac{1}{(t_s-t)^{2}}, \quad t\rightarrow t_s. \label{42}
\end{equation}
The entropy production thus diverges at the Big Rip, as we might expect. A noteworthy property of this expression is that $|\alpha|$, though present in the prefactor, is absent from the exponent. We see that $\dot{\sigma}(t)$ behaves similarly as the density $\rho(t)$, Equation~(\ref{36}), near the singularity.

\section{Summary} \label{s5}

A main lesson from this investigation is that the phantom barrier $w=-1$ should not be regarded as a sharp mathematical divide. The cosmic fluid is a complicated physical system, most probably endowed with properties such as viscosities such as common  elsewhere in fluid mechanics. Because of spatial symmetry, we have ignored the shear viscosity and maintained  the bulk viscosity only. As we have seen in Section \ref{s2}, the estimate~(\ref{23}) for the present-time bulk viscosity makes a passage through the phantom barrier quite possible in the late universe, even if it starts at present time $t=0$ from the quintessence region $w>-1 ~ (\alpha>0)$, at a maximum value $w=w_{\rm max}  ~(=0.056)$  inferred from the Planck data (2015) \cite{ade15}. Our comparison with experimental results was based mainly on the analysis of Wang and Meng~\cite{wang14}, although comparison with other analyses of Velten \emph{et al.} \cite{velten12,velten14} gave comparable results. This agreement is encouraging, and indicates that our numerical estimates are on the right track.

The qualitative derivations of the limiting forms of thermodynamic quantities  near the future singularity $t=t_s$ in Section \ref{s4}  were made on basis of the simplifying assumptions that both $w$ and the bulk viscosity $\zeta$ are constants. Near the singularity Equation~(\ref{42}) shows that the entropy production  diverges, as expected, whereas it could hardly have been seen beforehand that the exponent turns out to be independent of $|\alpha|$.

 It should be emphasized  that our theory is restricted to a one.component model. This is motivated, of course, by the dominant role of the dark fluid component. A more accurate description would be obtained by assuming $n$ components of the cosmic fluid,
\begin{equation}
\rho=\sum_{i=1}^n \rho_i, \quad P=\sum_{i=1}^n w_i\rho_i,
\end{equation}
($P$ being the total pressure, $w_i=$constants), we can insert these into the continuity equation
\begin{equation}
a\partial_a\rho(a)+3[\rho(a)+P]=3\zeta \theta,
\end{equation}
here expressed in terms of $a$. We intend to study this model more closely in Ref.~\cite{normann15}.

 As a closing remark,  it could be mentioned that the topic we have considered in this paper is related to the so-called quintom cosmology. The simplest quintom model is achieved by  introducing two scalar fields, one of them being quintessence and the other a phantom field. This kind of formulation arises from the desire to find alternative, and perhaps better descriptions, of the universe than inflationary theory. Crossing of the phantom barrier $w=-1$ is an important ingredient of the theory, and can even be described within the framework of braneworld cosmology. An extensive review of quintom cosmology is given by Cai \emph{et al.} \cite{cai10}. A recent paper relating quintom matter to the so-called emergent gravity can be found in Ref.~\cite{cai12}.

\newpage
\acknowledgments{Acknowledgments}
{I thank Ben David Normann for valuable information.}

\conflictofinterests{Conflicts of Interest}
{The author declares no conflict of interest.}

\bibliographystyle{mdpi}
\makeatletter
\renewcommand\@biblabel[1]{#1. }
\makeatother

\end{document}